\documentclass[12pt]{article}   
\pdfoutput=1
\usepackage{geometry,bbm}   
\usepackage{ulem}
\usepackage[toc,page]{appendix}             		
\usepackage{graphicx,cancel}
\usepackage[usenames,dvipsnames]{color}
\usepackage{slashed}					
\usepackage{amssymb,amsmath}
\usepackage{hyperref}
\usepackage{mathtools}
\usepackage{cite}

\usepackage{graphicx}
\usepackage{bm}
\def\bea{\begin{eqnarray}}
\def\eea{\end{eqnarray}}
\usepackage{latexsym}
\usepackage{epsfig,amssymb,euscript}
\usepackage{amsmath}
\usepackage{array,calc,epsfig}
\usepackage{bbold}

\newcommand{\be}{\begin{equation}}
\newcommand{\ee}{\end{equation}}

\numberwithin{equation}{section}
\include{macros}

\begin{document}

\begin{titlepage}

\begin{flushright}
SISSA 13/2019/FISI
\end{flushright}
\bigskip

\begin{center}
{\LARGE
{\bf

Charting the phase diagram of QCD$_\mathbf{3}$
}}
\end{center}

\bigskip
\begin{center}
{\large
Riccardo Argurio$^{1}$, Matteo Bertolini$^{2,3}$, Francesco Mignosa$^{2}$, Pierluigi Niro$^{1,4}$}
\end{center}

\renewcommand{\thefootnote}{\arabic{footnote}}

\begin{center}
\vspace{0.2cm}
$^1$ {Physique Th\'eorique et Math\'ematique and International Solvay Institutes, \\ Universit\'e Libre de Bruxelles; C.P. 231, 1050 Brussels, Belgium\\}
$^2$ {SISSA and INFN - 
Via Bonomea 265; I 34136 Trieste, Italy\\}
$^3$ {ICTP - 
Strada Costiera 11; I 34014 Trieste, Italy\\}
$^4$ {Theoretische Natuurkunde, Vrije Universiteit Brussel;\\ Pleinlaan 2, 1050 Brussels,
Belgium\\}
\vskip 5pt
{\texttt{rargurio@ulb.ac.be, bertmat@sissa.it, \\ fmignosa@sissa.it, pierluigi.niro@ulb.ac.be}}

\end{center}

\vskip 5pt
\noindent
\begin{center} {\bf Abstract} \end{center}
\noindent
We study the phase diagram of $SU(N)$ gauge theory in three space-time dimensions with a Chern-Simons term at level $k$, coupled to two sets of fundamental fermions with masses $m_1$ and $m_2$, respectively. The two-dimensional phase diagram that we propose shows a rich structure and widens in an interesting way previous results in the literature, to which it reduces in some limits. We present several checks of our proposal, including consistency with boson/fermion dualities. In this respect, we extensively comment on the structure of the scalar potential which is needed on the bosonic side of the duality.

\vspace{1.6 cm}
\vfill

\end{titlepage}

\newpage
\tableofcontents


\section{Introduction}
\label{intro}
Infrared (IR) dualities are useful tools to elucidate the non-perturbative dynamics of Quantum Field Theory (QFT). Usually, establishing these dualities requires technical tools that are available in two space-time dimensions or when supersymmetry is at work. However, in the last years progress in understanding the dynamics of gauge theories in three space-time dimensions has led to conjecture infinite families of non-supersymmetric dualities between different gauge theories coupled to (bosonic and/or fermionic) matter species \cite{Aharony:2015mjs,Hsin:2016blu,Metlitski:2016dht,Aharony:2016jvv,Komargodski:2017keh}. This originated from ideas coming from condensed matter physics \cite{Peskin:1977kp,Dasgupta:1981zz,Barkeshli:2014ida,Son:2015xqa,Wang:2015qmt,Potter:2015cdn,Seiberg:2016rsg,Wang:2016gqj,Seiberg:2016gmd,Murugan:2016zal} and large $N$ matter-coupled Chern-Simons theories \cite{Aharony:2011jz,Giombi:2011kc,Aharony:2012nh,GurAri:2012is,Aharony:2012ns,Jain:2013py,Jain:2013gza} and found several applications, including interesting connections with domain walls of four-dimensional QCD   \cite{Gaiotto:2017yup,Gaiotto:2017tne,Argurio:2018uup}. See {\it e.g.}  
 \cite{Karch:2016sxi,Karch:2016aux, Kachru:2016rui,Benini:2017dus,Jensen:2017dso,Gomis:2017ixy,Kachru:2016aon,Wang:2017txt,Jensen:2017xbs,Armoni:2017jkl,Cordova:2017vab,Benini:2017aed,Cordova:2017kue,Aharony:2018pjn,Choi:2018tuh, Sachdev:2018nbk,Benvenuti:2018cwd,Benvenuti:2019ujm, Aitken:2017nfd,Choudhury:2018iwf,Cordova:2018qvg,DiPietro:2019hqe} for more contributions in the  non-supersymmetric framework.
  
In this work we study the IR behavior of three-dimensional gauge theories in presence of a Chern-Simons term and coupled to Dirac fermions in the fundamental representation of the gauge group. In \cite{Komargodski:2017keh} the complete phase diagram of three-dimensional {$SU(N)_k$ gauge theory coupled to $F$ fundamental Dirac fermions with degenerate mass $m$ (one-family QCD$_3$, from now on) was analyzed as a function of $m$.\footnote{We adopt the usual convention where $G_k$ denotes a gauge group $G$ at level $k=k_{bare}-F/2$, with $k_{bare}\in\mathbb{Z}$ the bare Chern-Simons level of the UV gauge theory and $F$ the total number of fermions. Hereafter, we assume $k$ to be non-negative since one can always implement a time-reversal transformation, which maps $k$ into $-k$ and flips the sign of the fermion masses.}

We want to extend the analysis of \cite{Komargodski:2017keh} to the case in which the phase diagram has a two-dimensional structure. The simplest way to do this is to study an $SU(N)_k$ gauge theory coupled to $p$ fundamental fermions of mass $m_1$ and $F-p$ of mass $m_2$. Concretely, we are allowing mass deformations to explicitly break the global symmetry $U(F)$ to $U(p)\times U(F-p)$.\footnote{The global symmetry is actually given by the quotient $U(F)/\mathbb{Z}_N$, together with a discrete charge conjugation symmetry \cite{Benini:2017dus}. However, this does not affect our analysis and we will naively refer to the global symmetry as $U(F)$. 
} We will explore the phase diagram varying the two mass parameters and check that the dual bosonic theories, conjectured in \cite{Komargodski:2017keh}, admit the same phases as the fermionic theory, after deforming them with symmetry-breaking mass terms. Without loss of generality, we will consider $0\leq p\leq F/2$. 

On one hand, our analysis provides a non-trivial check that the conjectured boson/fermion dualities can be extended to more complicated cases and still maintain their validity. On the other hand, two-dimensional phase diagrams show a richer structure and novel phenomena with respect to one-family QCD$_3$ (including phase transitions between new gapless phases), and allow to perform some interesting vacuum analysis on the bosonic side of the duality, as well.

In the remainder of this section we briefly review the results of \cite{Komargodski:2017keh} 
about the phase diagram of one-family QCD$_3$. 
Then, in section \ref{phases}, we present our main results, {\it i.e.} our proposal for the  phase diagram when two different masses are varied independently.
Section \ref{checks} contains several checks of our proposal and gives also more details on the meaning of the different phases the theory enjoys  and of the phase transitions between them. We first focus on asymptotic phases, where one of the two masses is sent to $\pm \infty$. These asymptotic regions are effectively one-dimensional and the phase diagrams should reduce to those of QCD$_3$ with one species of matter fields. 
We then discuss in some detail the $k=0$ case, which is useful to perform some non-trivial consistency checks regarding time-reversal invariance and the Vafa-Witten theorem. We then analyze the vacua of the dual bosonic theories, conjectured from boson/fermion dualities, and show that they perfectly match the fermionic description in the neighborhoods of each  critical point.  Finally, we analyze some relevant deformations of the mass-degenerate sigma-model phase, which confirm, in a yet different way, our findings in a parametrically small neighborhood of the region with maximal global symmetry. Section \ref{conc} contains a discussion and an outlook.

\subsection{Review of one-dimensional phase diagram}
\label{KSreview}

The basic structure of the phase diagram of one-family QCD$_3$ analyzed in \cite{Komargodski:2017keh} is as follows.
\begin{enumerate}
\item If $k\geq F/2$ the global $U(F)$ symmetry is never broken and the theory has only two phases, which are visible semiclassically. For $m<0$ this is a topological $SU(N)_{k-\frac{F}{2}}$ phase, while for $m>0$ an $SU(N)_{k+\frac{F}{2}}$ one. This can be seen recalling that integrating out a massive fermion shifts the Chern-Simons level as $k \rightarrow k + \frac{1}{2}\, \text{sgn}(m)$.  Note that both phases are gapped, since the gauge bosons get a tree-level mass from the Chern-Simons term, but are non-trivial due to the topological nature of the Chern-Simons coupling. There exists a phase transition between these two different TQFTs, which, according to \cite{Aharony:2015mjs}, can be equivalently described by a bosonic dual theory.

Indeed, the two topological phases are level/rank dual\footnote{The precise form of level/rank dualities requires the presence of a gravitational Chern-Simons counterterm, see \textit{e.g.} \cite{Hsin:2016blu}. This term is generated on the $SU$ side of the duality once massive fermions are integrated out \cite{Seiberg:2016rsg}.} to $U(k-F/2)_{-N}$ and $U(k+F/2)_{-N}$. These are exactly the asymptotic phases of $U(k+F/2)_{-N}$ coupled to $F$ complex scalars in presence of a scalar potential. If one assumes that the $U(F)$ global symmetry is not broken and the dynamics prefers to maximally Higgs the gauge group, as the scalar mass $M^2$ is varied this bosonic theory has one single phase transition separating the same two phases as in the fermionic case. In fact, in appendix \ref{append1} we provide a proof of this vacuum pattern, starting from a scalar potential which contains all terms compatible with the symmetries of the problem, up to quartic order in the scalar fields. The phase diagram in the $k\geq F/2$ case is summarized in figure \ref{KS1}. 

\begin{figure}[t]
\centering
\includegraphics[scale=0.60]{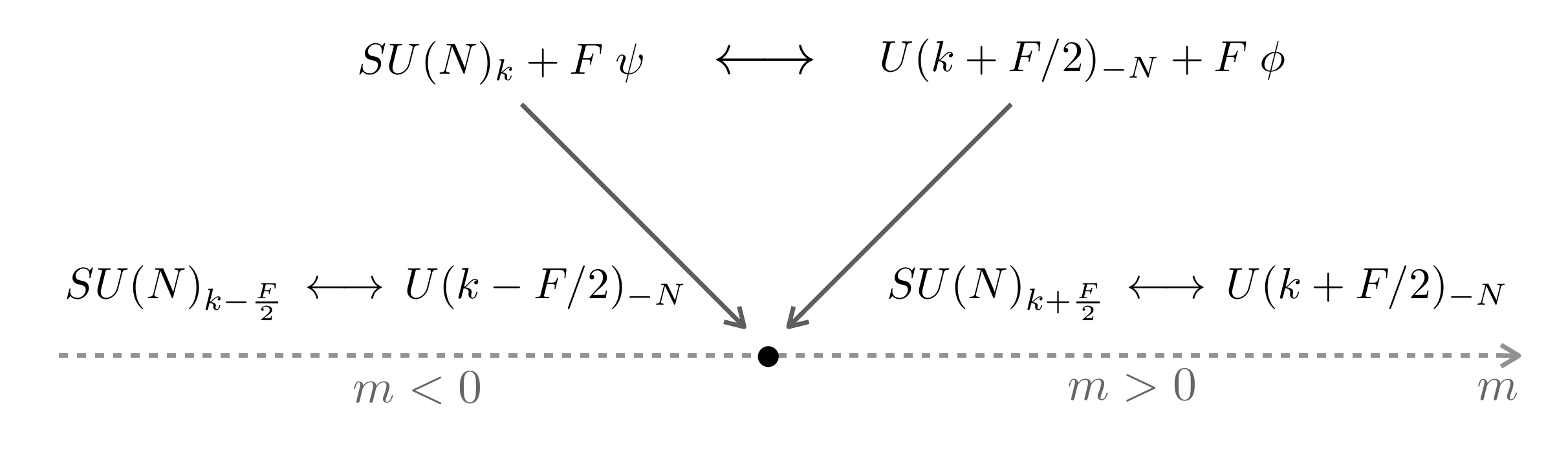}
	\caption{\small{Phase diagram of $SU(N)_k+F ~\psi$ in the case $k\geq F/2$. The phase transition between the two gapped phases is indicated by the black dot.}}
\label{KS1}
\end{figure}

For large values of $N$ and $k$ with fixed $k/N$, and for large $k$ and/or large $F$, it is known that the phase transition is unique and second order \cite{Appelquist:1988sr,Appelquist:1989tc,Aharony:2011jz,Giombi:2011kc,Aharony:2012nh,GurAri:2012is,Aharony:2012ns,Jain:2013py,Jain:2013gza}. In other regions of the parameter space, instead, the minimal structure of the phase diagram of \cite{Komargodski:2017keh} might change into a more intricate one. For example, recent results suggest that at large $N$ (with fixed $k$ and $F$) there are multiple first-order phase transitions \cite{Armoni:2019lgb}.\footnote{Note that even in this case, if one can match the vacuum structure on the two sides, boson/fermion dualities are still informative, despite being less powerful than in the case of second order phase transitions. We thank Zohar Komargodski for a discussion on this point.} Following \cite{Komargodski:2017keh}, we will assume that QCD$_3$ enjoys a single transition (though not necessarily second order) for values of $N, k$ and $F$ all not too large. 
  
\item If $k < F/2$ the previous picture is not true anymore,\footnote{Here and elsewhere, when considering low values of $k$, we assume $F$ to be smaller than the value $F^*$ above which the theory enjoys a single second-order phase transition (see \cite{Sharon:2018apk} for recent progress in rigorously estimating the bound $F^*$).} since the dual bosonic theory flows to a sigma model at low energies for negative $M^2$, being the global $U(F)$ symmetry spontaneously broken. The conjecture proposed in \cite{Komargodski:2017keh} is that the fermionic theory admits an inherently quantum phase where the quark bilinear condenses and breaks the global symmetry, leading to a sigma-model phase $\sigma$ with target space the complex Grassmannian
\be
\mbox{Gr}(F/2+k,F) = \frac{U(F)}{U(F/2+k)\times U(F/2-k)} ~,
\label{sigma}
\ee
and a Wess-Zumino term $\Gamma$ with coefficient $N$, see {\it e.g.} \cite{Witten:1983tw,Witten:1983tx,Freed:2017rlk}. Thus, in addition to the two asymptotic phases $SU(N)_{k-\frac{F}{2}}$ and $SU(N)_{k+\frac{F}{2}}$ (always visible semiclassically for large negative and large positive value of $m$, respectively), we also have a purely quantum sigma-model phase realized for small values of $|m|$ (of order of the gauge coupling constant $g^2$). Note that now the two asymptotic phases are level/rank dual to $U(F/2-k)_{N}$ and $U(k+F/2)_{-N}$, respectively. 

\begin{figure}[t]
	\centering
	\includegraphics[scale=0.62]{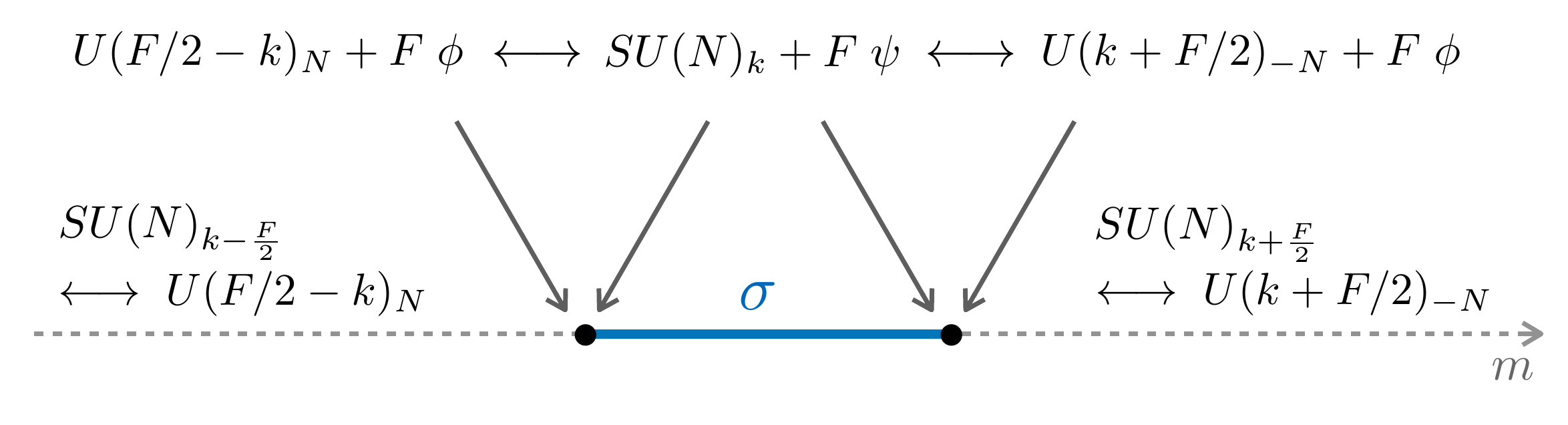}
	\caption{\small{Phase diagram of $SU(N)_k+F ~\psi$ in the case $k < F/2$. For $|m| \lesssim g^2$ the theory enters a quantum phase, with complex Grassmannian \eqref{sigma}. The two phase transitions, at negative and positive mass, are described by two different bosonic duals.}}
	\label{KS2}
\end{figure}

This means that in this case the phase diagram of the fermionic theory enjoys two phase transitions. 
These two transitions can be described by two different dual bosonic theories, $U(F/2-k)_{N}$ and $U(F/2+k)_{-N}$, each coupled to $F$ complex scalars. When the squared masses of the scalars in the two theories are large and positive, they flow respectively to the large negative and large positive mass phases of the fermionic theory. Instead, when the squared masses of the scalars are negative, both theories flow to the same sigma model with target space the complex Grassmannian (\ref{sigma}).  
As shown in appendix \ref{append1}, the same hypothesis on the structure of the scalar potential, which leads to maximal Higgsing for $k\geq F/2$, fixes, in the same negative squared mass regime, the sigma model to have target space (\ref{sigma}) in the present case.  

The phase diagram in the $k < F/2$ case is summarized in figure \ref{KS2}. 

\end{enumerate}
We now present our proposal on how the one-dimensional phase diagrams discussed above get extended when allowing for symmetry-breaking mass deformations.

\section{The two-dimensional phase diagram}
\label{phases}

The phase diagram of $SU(N)_k$ gauge theory coupled to  $p$ fundamental fermions $\psi_1$ and $F-p$ fundamental fermions $\psi_2$, as a function of $m_1$ and $m_2$, turns out to have a different structure, depending on the value of the Chern-Simons level $k$ at fixed $F$ and $p$: $k \geq F/2$, $F/2-p\leq k < F/2$ and $0\leq k <F/2-p$. 

A property of all phase diagrams is that on the diagonal line $m_1=m_2$, where the global symmetry is enhanced to the full $U(F)$, one correctly recovers the corresponding one-dimensional diagram reviewed in previous section. We now illustrate the three different phase diagrams in turn.

\subsection*{$k \geq F/2$}

In this case the phase diagram, which is shown in figure \ref{kbig}, presents only phases which are visible semiclassically. Consistently, as we are going to show in section \ref{bosondual}, the full phase space can be equivalently described in terms of a {\it unique} dual bosonic theory, with gauge group $U(k+F/2)_{-N}$ and two sets of $p$ and $F-p$ complex scalars in the fundamental representation. 

Conventions are as follows. The black dot at the origin represents the usual phase transition of the   $SU(N)_k + F~\psi$ theory. Perturbing it by two independent mass deformations, proportional to $m_1$ and $m_2$, one covers a two-dimensional space, enjoying four different topological phases $\mbox{T}_i$ defined as
\begin{alignat}{3}
\mbox{T}_1 &: SU(N)_{k+ \frac{F}{2}} &~\longleftrightarrow ~ &U(k+F/2)_{- N} \\
\mbox{T}_2 &: SU(N)_{k+ \frac{F}{2}-p} &~\longleftrightarrow ~ &U(k+F/2-p)_{- N} \\
\mbox{T}_3 &: SU(N)_{k- \frac{F}{2}} &~\longleftrightarrow ~ &U(k-F/2)_{- N} \\
\mbox{T}_4 &: SU(N)_{k- \frac{F}{2}+p} &~\longleftrightarrow ~ &U(k-F/2+p)_{- N} 
\end{alignat}
where $\longleftrightarrow$ stands for level/rank duality. Note that in the limiting case $k=F/2$, the topological theory T$_3$ becomes trivially gapped.
Note also that, consistently, T$_1$ and T$_3$ are the same topological phases one expects for the theory with common mass $m=m_1=m_2$ in the range $k \geq F/2$ (cf figure \ref{KS1}) for positive and negative $m$, respectively, and  which one should recover on the bisector of the first and third quadrants of figure \ref{kbig}. 

Red lines represent phase transitions in the $(m_1,m_2)$ plane which are absent in the one-dimensional phase diagrams. For instance, each point on the red line separating phases T$_1$ and T$_2$ defines a critical theory $SU(N)_{k+\frac{F}{2}-\frac{p}{2}}+p~\psi_1$. In bosonic language, this can be equivalently described by $U(k+F/2)_{-N}+p~\phi_1$. The same logic applies to all other red lines. Consistency with boson/fermion duality, which we elaborate upon in section \ref{bosondual}, suggests that the four red lines do indeed meet at a single point (black dot in the figure).

\begin{figure}[t]
\centering
\includegraphics[scale=0.52]{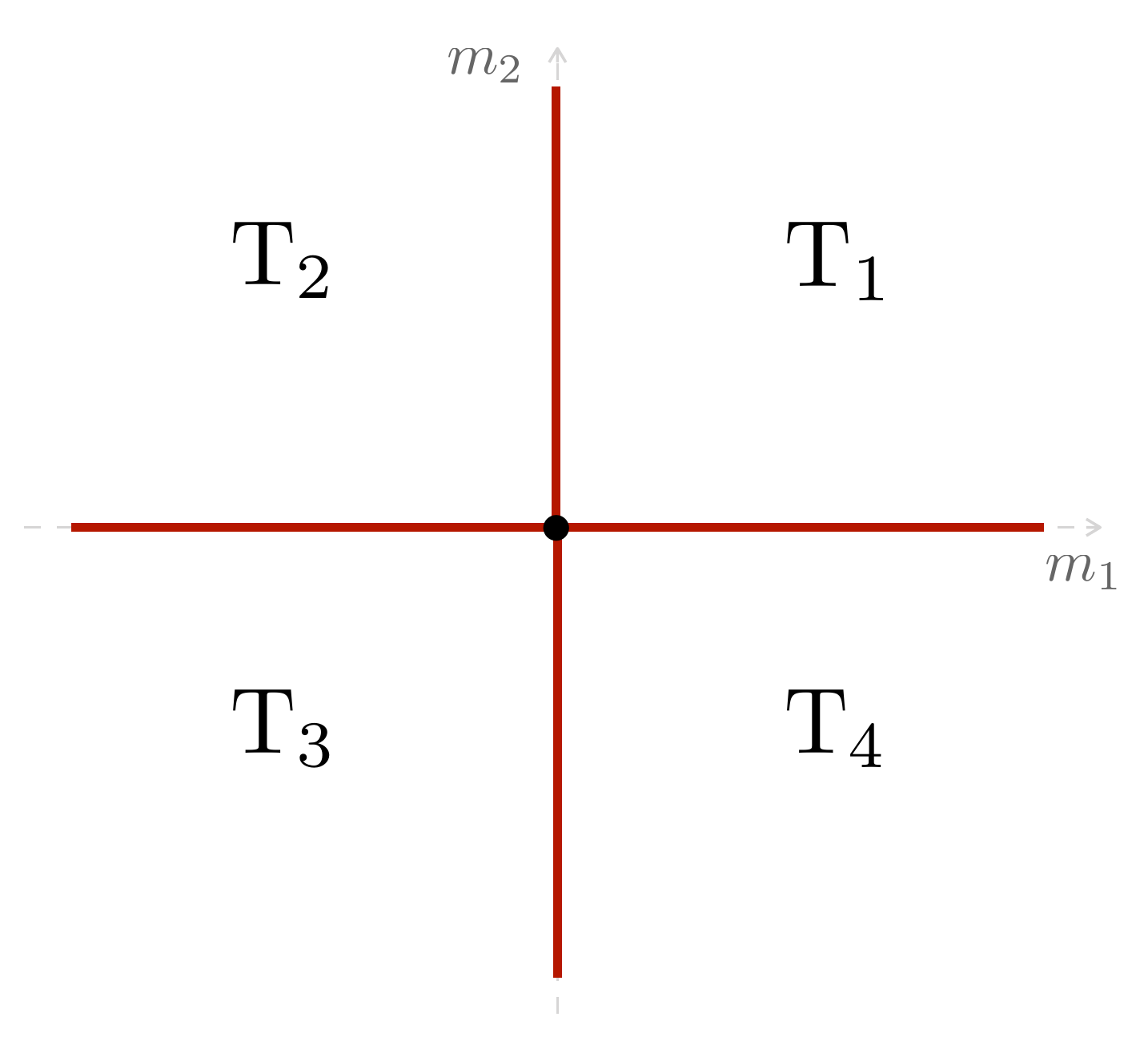}
	\caption{\small{Phase diagram of $SU(N)_k + p~\psi_1+(F-p)~\psi_2$ in the case $k \geq F/2$.}}
\label{kbig}
\end{figure}

\subsection*{$F/2-p \leq k < F/2$}

In this case, besides genuine topological phases, the phase diagram presents three inherently quantum phases. This implies, as we show explicitly in section \ref{bosondual}, that two different dual bosonic descriptions are needed to cover the full fermionic phase diagram, {\it i.e.} $U(F/2-k)_N+p~\phi_1+(F-p)~\phi_2$ and $U(F/2+k)_{-N}+p~\phi_1+(F-p)~\phi_2$. 

The phase diagram is reported in figure \ref{kmedium}. 
\begin{figure}[t]
\centering
\includegraphics[scale=0.52]{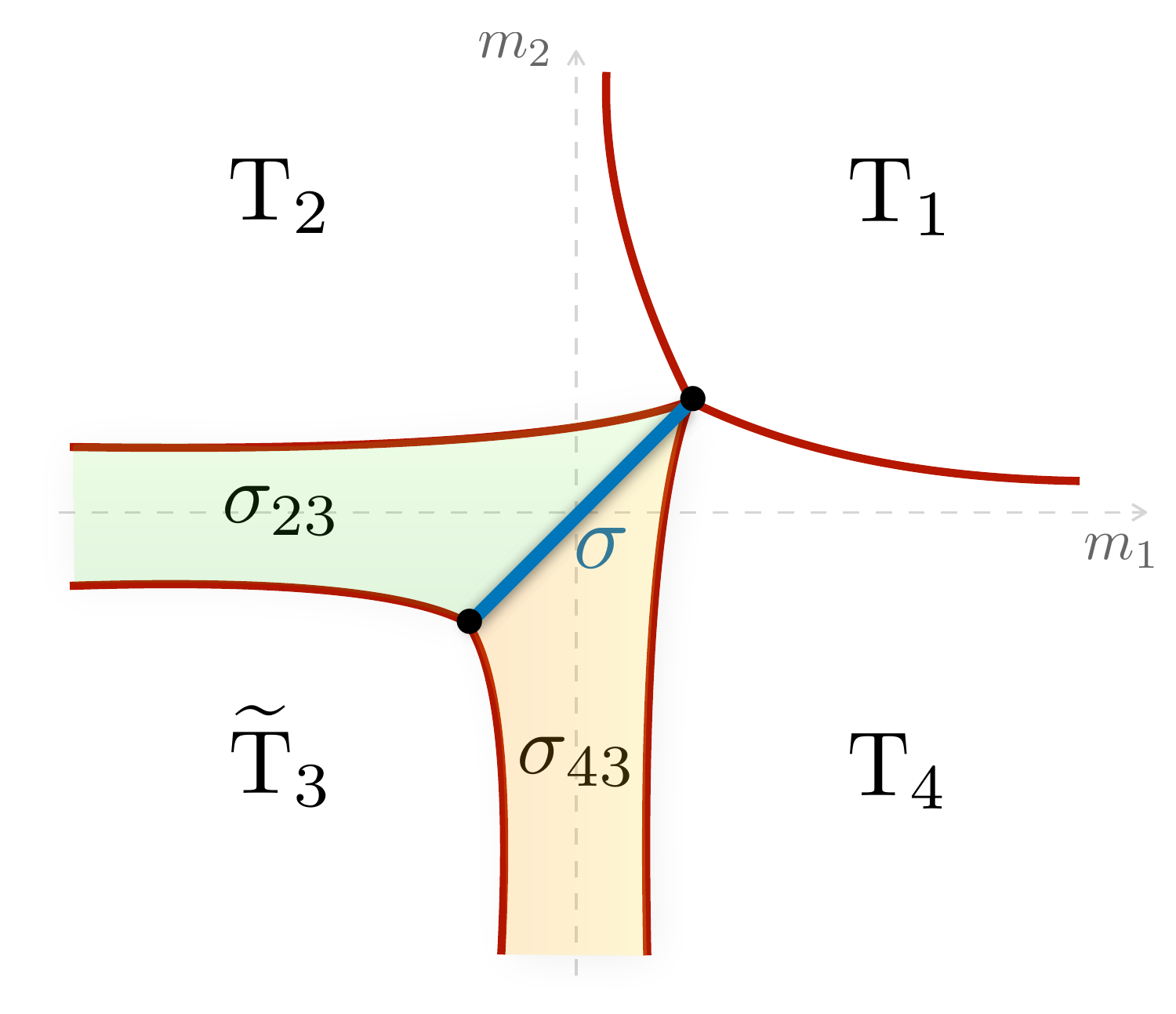}
	\caption{\small{Phase diagram of $SU(N)_k + p~\psi_1+(F-p)~\psi_2$ in the case $F/2-p \leq k < F/2$.}}
\label{kmedium}
\end{figure}
The black dots represent the two phase transitions of the degenerate mass case, $m_1=m_2$, cf figure \ref{KS2}. The topological theories are now 
\begin{alignat}{3}
\mbox{T}_1 &: SU(N)_{k+ \frac{F}{2}} &~\longleftrightarrow ~ &U(k+F/2)_{- N} \\
\mbox{T}_2 &: SU(N)_{k+ \frac{F}{2}-p} &~\longleftrightarrow ~ &U(k+F/2-p)_{- N} \\
\widetilde{\mbox{T}}_3 &: SU(N)_{k- \frac{F}{2}} &~\longleftrightarrow ~ &U(F/2-k)_{N} \\
\mbox{T}_4 &: SU(N)_{k- \frac{F}{2}+p} &~\longleftrightarrow ~ &U(k-F/2+p)_{- N} 
\end{alignat}
where, again, T$_1$ and $\widetilde{\mbox T}_3$ are the correct topological phases one should find on the bisector, cf figure \ref{KS2}. The blue line represents a quantum phase, with target space \eqref{sigma} and a Wess-Zumino term with coefficient $N$. Finally, the shaded regions in the plane refer to sigma-model phases, where the IR dynamics is not gapped, but it is described by  non-linear sigma models with different target spaces and a Wess-Zumino term with coefficient $N$. In particular, the target space of $\sigma_{23}$ is the complex Grassmannian 
\be
\label{g1}
\mbox{Gr}(F/2-k,F-p)=\frac{U(F-p)}{U(F/2-k)\times U(k+F/2-p)} ~,
\ee 
and that of $\sigma_{43}$ is 
\be
\label{g2}
\mbox{Gr}(F/2-k,p)=\frac{U(p)}{U(F/2-k)\times U(k-F/2+p)}~.
\ee
In the limiting case $k=F/2$ all sigma-model phases $\sigma_{23}$, $\sigma_{43}$ and $\sigma$ trivialize, as well as the topological phase $\widetilde{\mbox{T}}_3$. Thus, the two phase diagrams in figures \ref{kbig} and \ref{kmedium} become topologically equivalent, as expected. In the other limiting case, $k=F/2-p$, to which we connect next, it is the sigma-model phase $\sigma_{43}$ and the topological phase $\mbox{T}_4$ which trivialize, instead.

As in figure \ref{kbig}, red lines represent phase transitions in the $(m_1,m_2)$ plane. Here, however, there also exist lines separating topological and massless phases. For instance, each point on the red line separating T$_2$ and $\sigma_{23}$, and the corresponding one separating the latter with $\widetilde{\mbox{T}}_3$, are the two phase transitions one expects for $SU(N)_{k-\frac{p}{2}}+(F-p)~\psi_2$. According to \cite{Komargodski:2017keh}, these two phase transitions are described by two different bosonic duals, $U(k+F/2-p)_{-N}+(F-p)~\phi_2$ and $U(F/2-k)_{N}+(F-p)~\phi_2$, respectively (similar arguments hold when looking at $\sigma_{43}$ as the gapless phase  separating the topological phases T$_4$ and $\widetilde{\mbox{T}}_3$, the relevant fermionic theory on the red lines being now $SU(N)_{k-\frac{F}{2}+\frac{p}{2}}+p~\psi_1$).

Note that not all lines cutting through the two-dimensional phase diagram can be effectively reduced to a one-dimensional phase diagram of a single family theory.
This applies, in particular, to the region of small masses, where different gapless quantum phases meet and $\sigma$ becomes a phase transition itself, which separates the gapless phases $\sigma_{23}$ and $\sigma_{43}$. This is a novel phenomenon, which does not have any counterpart in one-family QCD$_3$. Indeed, in our case the pattern of symmetry breaking is richer, giving a variety of quantum phases which meet in the region where both masses are, in modulus, $\lesssim g^2$. 

\subsection*{$0\leq k<F/2-p$}
Also in this range of parameters the phase diagram presents three different quantum phases. The two dual bosonic descriptions needed to cover the full phase space are again 
$U(F/2-k)_N+p~\phi_1+(F-p)~\phi_2$ and $U(F/2+k)_{-N}+p~\phi_1+(F-p)~\phi_2$. The phase diagram, where the same conventions as before are adopted, is shown in figure \ref{ksmall}. 

\begin{figure}[t]
\centering
\includegraphics[scale=0.52]{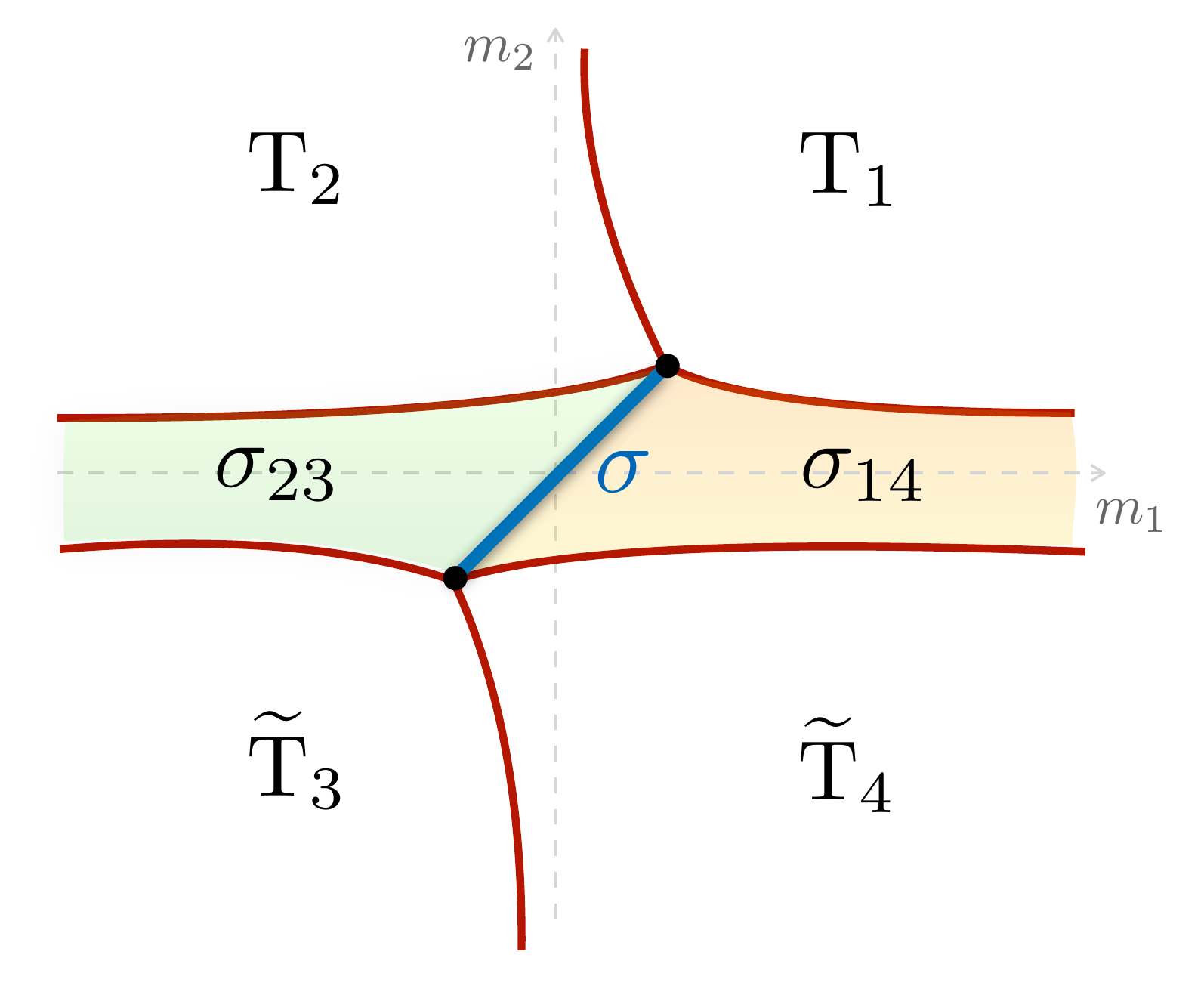}
	\caption{\small{Phase diagram of $SU(N)_k + p~\psi_1+(F-p)~\psi_2$ in the case $0 \leq k < F/2-p$.}}
\label{ksmall}
\end{figure}

The topological phases are now 
\begin{alignat}{3}
\mbox{T}_1 &: SU(N)_{k+ \frac{F}{2}} &~\longleftrightarrow ~ &U(k+F/2)_{- N} \\
\mbox{T}_2 &: SU(N)_{k+ \frac{F}{2}-p} &~\longleftrightarrow ~ &U(k+F/2-p)_{- N} \\
\widetilde{\mbox{T}}_3 &: SU(N)_{k- \frac{F}{2}} &~\longleftrightarrow ~ &U(F/2-k)_{N} \\
\widetilde{\mbox{T}}_4 &: SU(N)_{k- \frac{F}{2}+p} &~\longleftrightarrow ~ &U(F/2-p-k)_{N} 
\end{alignat}
The sigma-model phases $\sigma$ and $\sigma_{23}$ are as before while $\sigma_{14}$ has target space the complex Grassmannian 
\be
\label{g3}
\mbox{Gr}(F/2+k,F-p)=\frac{U(F-p)}{U(F/2+k)\times U(F/2-p-k)}~,
\ee 
and a Wess-Zumino term with coefficient $N$.

In the limiting case $k=F/2-p$, the phase $\sigma_{14}$ and the topological phase $\widetilde{\mbox{T}}_4$ trivialize. For $k=F/2-p$, the two phase diagrams in figure \ref{kmedium} and \ref{ksmall} become thus topologically equivalent, as expected. Again, the quantum phase $\sigma$ separates two different phases described by the two Grassmannians \eqref{g1} and \eqref{g3}. Note, in particular, that for sufficiently low values of $|m_2|$, the theory enjoys only sigma-model phases for all values of $m_1$ (the asymmetry between $m_1$ and $m_2$ is due to our choice $p \leq F/2$).

Let us close this discussion considering a few specific values for $p$. 
 
When $p=0$ the phase diagram of figure \ref{kmedium} disappears since its allowed range for $k$ becomes an empty set. Figures  \ref{kbig} and \ref{ksmall}, instead, collapse to a single vertical line, the $m_2$ axis, and their topology becomes the same as the one-dimensional diagrams of figures  \ref{KS1} and \ref{KS2}, respectively, as one should clearly expect. In particular, the sigma models $\sigma$, $\sigma_{14}$ and $\sigma_{23}$ become identical for $p=0$, while $\mbox{T}_1=\mbox{T}_2$, $\mbox{T}_3=\mbox{T}_4$ and $\widetilde{\mbox{T}}_3=\widetilde{\mbox{T}}_4$.

When $p=F/2$, the phase diagrams in figure \ref{kbig} and \ref{kmedium} should be symmetric with respect to the $m_1=m_2$ line, while it is the phase diagram in figure \ref{ksmall}  which now disappears. Consistently, the sigma models with target spaces $\sigma_{23}$ and $\sigma_{43}$ coincide for $p=F/2$, as well as the topological phases $\mbox{T}_2$ and $\mbox{T}_4$.

\section{Consistency checks and beyond}
\label{checks}

We now present various checks for the validity of our proposed two-dimensional phase diagrams.

\subsection{Asymptotic phases: matching ordinary QCD$_\mathbf{3}$}
\label{asympt}

The proposed two-dimensional phase diagrams should satisfy various consistency checks with the results in \cite{Komargodski:2017keh}. The simplest one is that the phase diagram of one-family QCD$_3$ should be recovered on the $m_1=m_2$ line, where the global symmetry is enhanced to $U(F)$. This is something we have already noticed to hold. A more intricate set of checks comes by studying extreme mass regimes. 

Starting from the original $SU(N)_k + p~\psi_1+(F-p)~\psi_2$ theory, let us consider the four different theories one obtains by integrating out, with either signs for the mass, one of the two fermion families, $\psi_1$ and $\psi_2$. These are $SU(N)$ gauge theories with a shifted Chern-Simons level and coupled to $p$ or $F-p$ fundamental fermions with mass $m_1$ or $m_2$, respectively. The IR phases of these theories are easily constructed by the same methods of \cite{Komargodski:2017keh}. Such phases should coincide with the ones of our  two-dimensional diagrams, figures \ref{kbig}, \ref{kmedium} and \ref{ksmall}, in the asymptotic, large mass regions. The four asymptotic theories and their one-dimensional phase diagrams are the following:
\begin{enumerate}
\item $m_1 \rightarrow + \infty $: one ends up with $SU(N)_{k+ \frac{p}{2}} + (F-p) ~ \psi_2$, which has only the two semiclassical phases if $k\geq F/2-p$. Its phase diagram has the following structure:
\begin{itemize}
\item $k\geq F/2-p$: the two phases are $\mbox{T}_1$ (for positive $m_2$) and $\mbox{T}_4$ (for negative $m_2$). The dual bosonic theory is $U(k+F/2)_{-N}+(F-p)~\phi_2$.
\item $0\leq k < F/2-p$: the topological phases are $\mbox{T}_1$ (for positive $m_2$) and $\widetilde{\mbox{T}}_4$ (for negative $m_2$), while the intermediate sigma-model phase is $\sigma_{14}$. The dual bosonic theories are $U(k+F/2)_{-N}+(F-p)~\phi_2$ (for positive $m_2$) and $U(F/2-k-p)_{N}+(F-p)~\phi_2$ (for negative $m_2$).
\end{itemize}

\item $m_1  \rightarrow - \infty$: one gets $SU(N)_{k-\frac{p}{2}} + (F-p) ~ \psi_2$, which has only the two semiclassical phases if $k\geq F/2$. Its phase diagram has the following structure:
\begin{itemize}
\item $k\geq F/2$: the two phases are $\mbox{T}_2$ (for positive $m_2$) and $\mbox{T}_3$ (for negative $m_2$). The dual bosonic theory is $U(k+F/2-p)_{-N}+(F-p)~\phi_2$.
\item $0\leq k < F/2$: the topological phases are $\mbox{T}_2$ (for large positive $m_2$) and $\widetilde{\mbox{T}}_3$ (for large negative $m_2$), while the intermediate sigma-model phase is $\sigma_{23}$. The dual bosonic theories are $U(k+F/2-p)_{-N}+(F-p)~\phi_2$ (for positive $m_2$) and $U(F/2-k)_{N}+(F-p)~\phi_2$ (for negative $m_2$).
\end{itemize}

\item $m_2  \rightarrow + \infty$: one ends up with $SU(N)_{k+\frac{F}{2}-\frac{p}{2}} + p ~ \psi_1$, which has only two semiclassical phases for any non-negative $k$. The two phases are $\mbox{T}_1$ (for positive $m_1$) and $\mbox{T}_2$ (for negative $m_1$). The dual bosonic theory is $U(k+F/2)_{-N}+p~\phi_1$.

\item $m_2  \rightarrow - \infty$: one gets $SU(N)_{k-\frac{F}{2}+\frac{p}{2}} + p ~ \psi_1$, which has only two semiclassical phases if $k\geq F/2$ or $0 \leq k\leq F/2-p$. Its phase diagram has the following structure:
\begin{itemize}
\item $k\geq F/2$: the two phases are $\mbox{T}_4$ (for positive $m_1$) and $\mbox{T}_3$ (for negative $m_1$). The dual bosonic theory is $U(k-F/2+p)_{-N}+p~\phi_1$.
\item $F/2-p\leq k < F/2$: the topological phases are $\mbox{T}_4$ (for positive $m_1$) and $\widetilde{\mbox{T}}_3$ (for negative $m_1$), and the intermediate sigma-model phase is $\sigma_{43}$. The dual bosonic theories are $U(k-F/2+p)_{-N}+p~\phi_1$ (for positive $m_1$) and $U(F/2-k)_{N}+p~\phi_1$ (for negative $m_1$).
\item $0\leq k < F/2-p$: the two phases are $\widetilde{\mbox{T}}_4$ (for positive $m_1$) and $\widetilde{\mbox{T}}_3$ (for negative $m_1$). The dual bosonic theory is $U(F/2-k)_{N}+p~\phi_1$.
\end{itemize}
\end{enumerate}
It is easy to check that our proposed phase diagrams, figures \ref{kbig}, \ref{kmedium} and \ref{ksmall}, exactly reproduce this intricate structure in the large $|m_1|$ and/or $|m_2|$ regions.

\begin{figure}[t]
\centering
\includegraphics[scale=0.5]{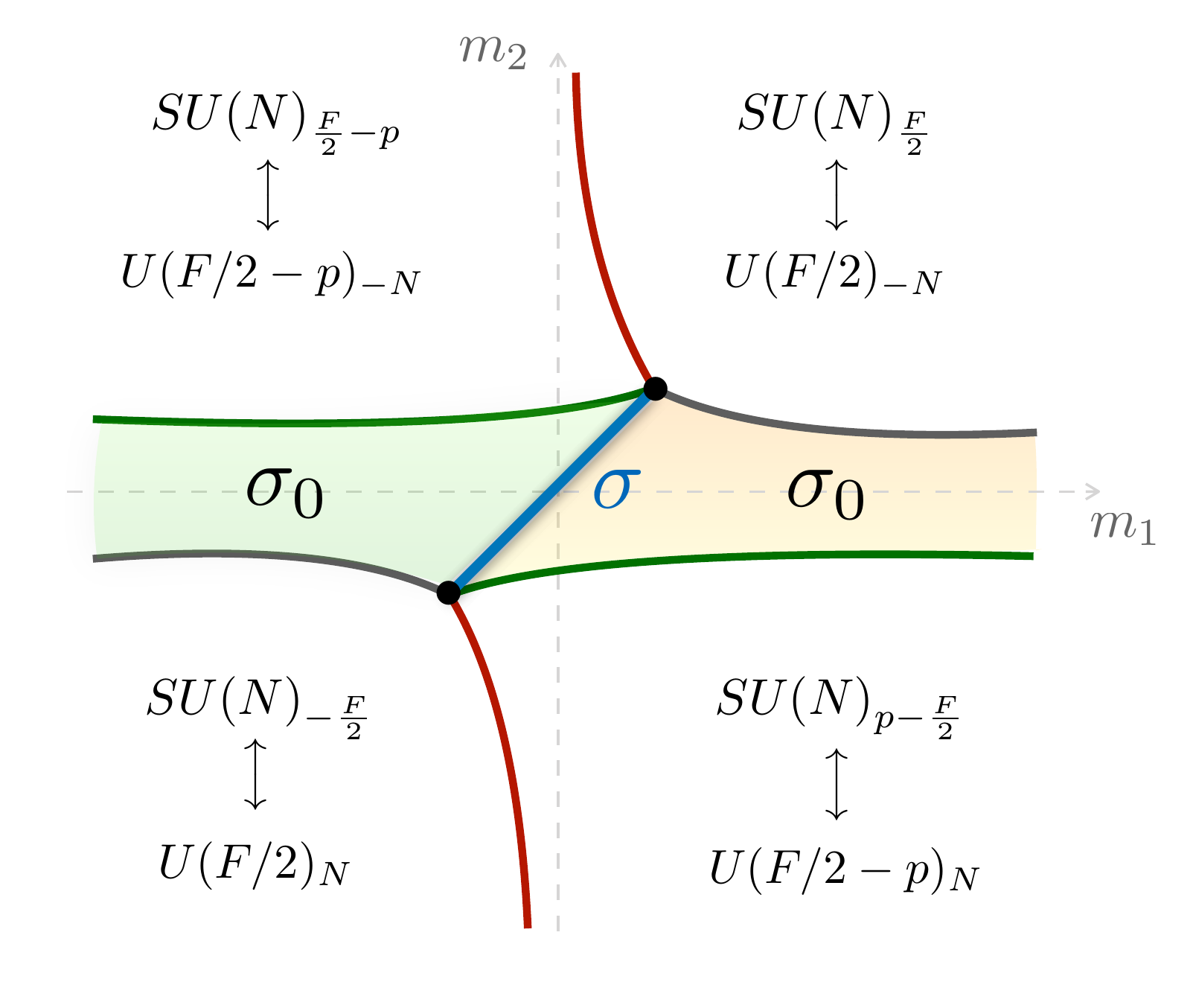}
	\caption{\small{Phase diagram of $SU(N)_0 + p~\psi_1+(F-p)~\psi_2$. The transition lines with the same color are described by the same theories, up to a time-reversal transformation.}}
\label{k0gen}
\end{figure}


\subsection{$k=0:$ time reversal and Vafa-Witten theorem}
\label{kappa0}

One interesting non-trivial check comes by taking $k=0$, in which the theory we study becomes $SU(N)_0+p~\psi_1+(F-p)~\psi_2$. Since $k=0$, time-reversal acts on this theory just flipping the sign of the mass terms of the two sets of fermions. As a consequence, the two-dimensional phase diagram should be symmetric with respect to the origin, modulo the flipping of the effective Chern-Simons levels of the specular phases. This symmetry can be nicely observed in  the phase diagram in figure \ref{ksmall}, which we report in figure \ref{k0gen} for the particular case $k=0$.

\begin{figure}[t]
\centering
\includegraphics[scale=0.5]{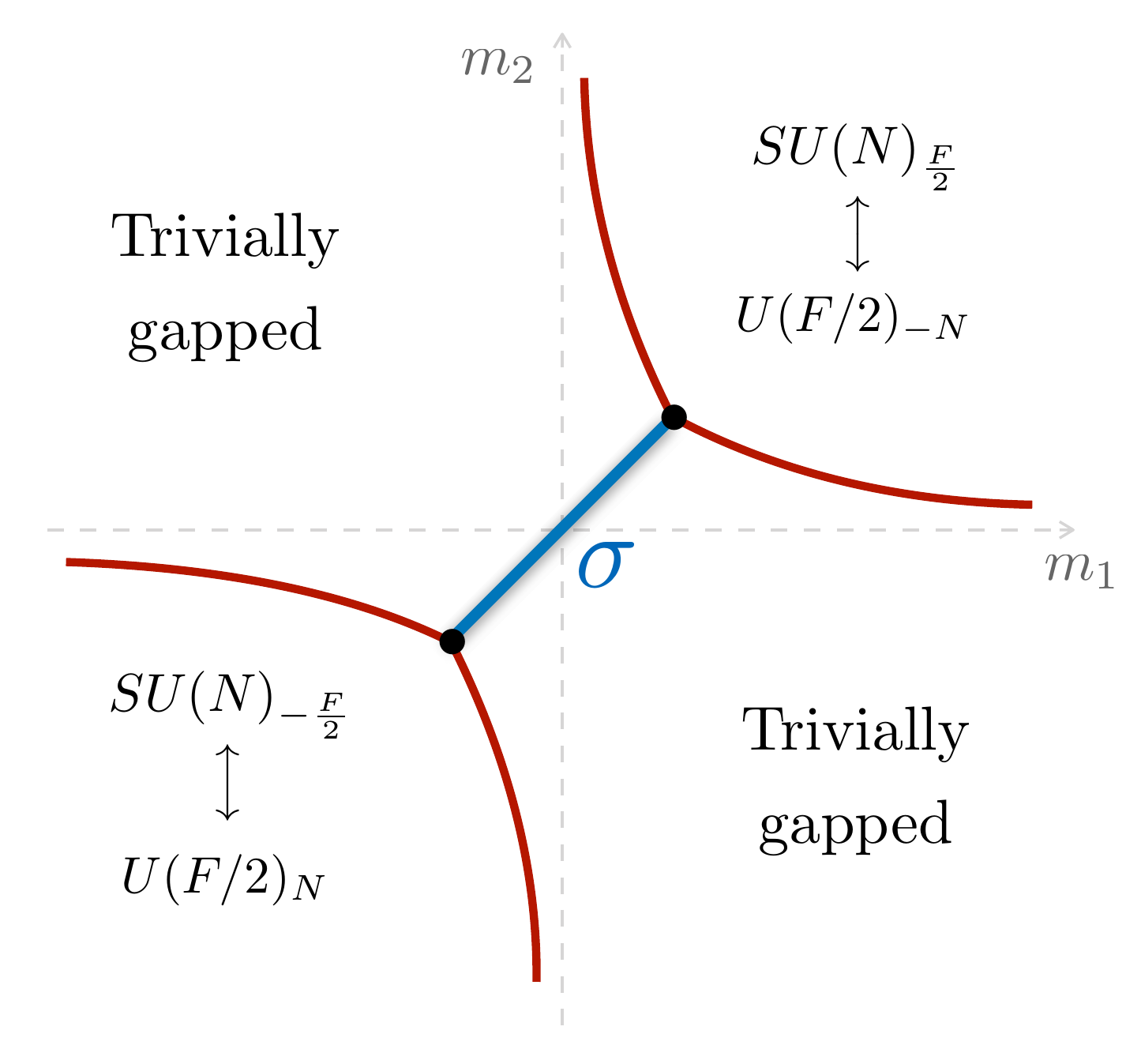}
	\caption{\small{Phase diagram of $SU(N)_0 + F/2~\psi_1+F/2~\psi_2$. In the bosonic dual picture, the red transition lines in the first and third quadrant are described by $U(F/2)_{\mp N}+F/2~\phi$, respectively. On the two black spots, where the global symmetry is enhanced, the number of scalars is $F$.}}
\label{k0pF2}
\end{figure}

It is a further consistency check that the sigma models with target spaces (\ref{g1}) and (\ref{g3}) coincide for $k=0$. We labelled this sigma model with $\sigma_0$, whose target space is
\be
\mbox{Gr}(F/2,F-p) = \frac{U(F-p)}{U(F/2)\times U(F/2-p)} ~,
\label{sigmapm}
\ee
and which includes a Wess-Zumino term with coefficient $N$. The sigma-model phases $\sigma_0$ and $\sigma$ enjoy time-reversal invariance for $k=0$, as it should be \cite{Komargodski:2017keh}.

In addition to $k=0$, let us also take $p=F/2$, {\it i.e.} we consider the same number of fermions in the two sets. In this case the phase diagram should be also symmetric with respect to the $m_1=m_2$ line, besides being symmetric with respect to the origin. The phase diagram for $k=0$ and $p=F/2$ is depicted in figure \ref{k0pF2}. We easily see that $\sigma_0$ trivializes, as well as the topological theories $SU(N)_{p-\frac{F}{2}}$ and $SU(N)_{\frac{F}{2}-p}$, leaving a trivially gapped phase in the second and fourth quadrants. Moreover, the theories on the two red curves are the same, up to the sign of the Chern-Simons level.

Another interesting check when $k=0$ and $p=F/2$ is related to the expected enhancement of time-reversal symmetry on the full $m_2=-m_1$ line. This one-dimensional slice of the diagram of figure \ref{k0pF2} can be interpreted as a smooth gapping of an original $SU(N)_0$ coupled to $F$ massless fundamental fermions, which is time-reversal invariant and has been suggested to break the full $U(F)$ global symmetry to $U(F/2) \times U(F/2)$ at strong coupling. Then, the Vafa-Witten theorem \cite{Vafa:1984xg,Vafa:1984xh,Vafa:1983tf} prevents this theory from developing a further symmetry breaking of flavor and time-reversal symmetry along the line $m_2=-m_1$. As a consequence, on such line  fermions can be safely integrated out, leading to a trivially gapped vacuum outside the origin. At the origin, the breaking $U(F) \rightarrow U(F/2) \times U(F/2)$ gives rise to the  target space \eqref{sigma} with $k=0$. The diagram in figure \ref{k0pF2} exactly reproduces all these features.

\subsection{Dual bosonic theories: matching the phase transitions}
\label{bosondual}

In section \ref{asympt} we have checked our two-dimensional phase diagrams in the large mass regime, where they become effectively one-dimensional, against one-family $\mbox{QCD}_3$. Here we want to focus on the region near the critical points, \textit{i.e.} the black dots in figures \ref{kbig}, \ref{kmedium} and \ref{ksmall}. This is done using  boson/fermion duality, properly adapted to the two-family case. This will also work as a nice consistency check of the duality itself. 

From the conjectured boson/fermion duality of the one-family case, one can argue that the bosonic theories one should consider near the critical points are 
\be
U(n)_l+p~\phi_1+(F-p)~\phi_2 ~,
\label{bosongeneric}
\ee
where $(n,l)=(F/2\pm k,\mp N)$ and $\phi_1,\phi_2$ are scalar fields in the fundamental representation of the gauge group. Here, we have explicitly split the $F$ scalars in two different sets, since we want to deform the massless theories describing the critical points with the independent massive deformations, $M^{\,2}_1$ and $M^{\, 2}_2$ respectively.

We can reproduce the desired vacuum structure assuming that, when at least one of the two sets condenses, the gauge group is maximally Higgsed and the unbroken global symmetry is maximized. In fact, exactly as for the one-family model discussed in appendix \ref{append1}, it is possible to show that these assumptions hold true if we consider the scalar potential of the critical theory up to quartic order in the scalar fields, and then deform it with symmetry-breaking mass operators. In terms of the gauge invariant operators $X=\phi_1 \phi_1^\dagger$, $Y=\phi_2 \phi_2^\dagger$ and $Z=\phi_1 \phi_2^\dagger$, we can write the (deformed) potential as
\be
V= M^{\, 2}_1 \mbox{Tr}X + M^{\, 2}_2 \mbox{Tr}Y + \lambda (\mbox{Tr}^2 X + \mbox{Tr}^2 Y + 2\mbox{Tr}X\mbox{Tr}Y) \\
+ \widetilde\lambda ( \mbox{Tr}X^2 + \mbox{Tr}Y^2 + 2\mbox{Tr}Z Z^{\dagger}) ~,
\ee
where $X$ and $Y$ are positive semidefinite Hermitian matrices of dimension $p$ and $F-p$, respectively, whereas $Z$ is a $p\times F-p$ rectangular matrix. 
Note that the quartic couplings in the potential are chosen to respect the full $U(F)$ symmetry. This is because we are limiting ourselves to perturbations due to massive deformations only. In principle, there could be other $U(p) \times U(F-p)$ preserving relevant deformations besides massive ones. If boson/fermion duality is correct, these deformations should have a counterpart on the fermionic side, but we do not consider them here.

For the same reasons as the one-family case discussed in appendix \ref{append1}, we take $\widetilde{\lambda}>0$, which requires $\widetilde{\lambda} + \mbox{min}(n,F) \lambda > 0$ in order for the potential to be bounded from below. 

In the first quadrant of the $(M^{\,2}_1,M^{\, 2}_2)$ plane, where both $M^{\, 2}_1$ and $M^{\,2}_2$ are positive, $X$, $Y$ and $Z$ vanish on shell. This implies that there is no scalar condensation, all matter fields are massive and can be integrated out, leading to a $U(n)_l$ topological theory in the IR.

In all other cases, the vacuum equations imply that $Z=0$, while $X$ and $Y$ are diagonal with respectively $r_1$ and $r_2$ degenerate non-negative eigenvalues given by
\be
\begin{split}
x&=\frac{-(\widetilde\lambda+\lambda r_2)M^{\, 2}_1+\lambda r_2 M^{\, 2}_2}{2\widetilde\lambda^2+2\lambda\widetilde\lambda(r_1+r_2)} ~, \\
y&=\frac{-(\widetilde\lambda+\lambda r_1)M^{\, 2}_2+\lambda r_1 M^{\, 2}_1}{2\widetilde\lambda^2+2\lambda\widetilde\lambda(r_1+r_2)} ~. \\
\end{split}
\ee
The positivity condition on $x$ and $y$ implies that a simultaneous condensation of both $\phi_1$ and $\phi_2$ is only allowed in a subregion $\cal R$ of the third quadrant of the ($M^{\, 2}_1$,$M^{\, 2}_2$) plane, which includes the line $M_1^{\,2}=M_2^{\, 2}$. Outside this region and above (below) the bisector, only $x$ ($y$) can be non-zero, meaning that only $\phi_1$ ($\phi_2$) can condense. The ranks $r_1$ and $r_2$ are non-negative integers which satisfy the constraints
\be
\begin{split}
0 \leq r_1 \leq \mbox{min}(n,p) ~~,~~0 \leq r_2 \leq \mbox{min}(n,F-p) ~~,~~r_1+r_2 \leq \mbox{min}(n,F) ~,
\end{split}
\label{constraint}
\ee
and have to be determined by minimizing the vacuum potential, seen as a function of ($r_1,r_2$). Once we determine these values, the spontaneous symmetry breaking of the flavor symmetry follows the pattern
\be
U(p)\times U(F-p) \longrightarrow U(r_1)\times U(p-r_1)\times U(r_2)\times U(F-p-r_2) ~.
\ee
This leads, in the region where $x$ and $y$ do not vanish simultaneously, to a sigma model with coset
\be
\mathrm{Gr}(r_1,p)\times\mathrm{Gr}(r_2,F-p) =\frac{U(p)}{U(r_1)\times U(p-r_1)} \times \frac{U(F-p)}{U(r_2)\times U(F-p-r_2)} ~,
\label{sigmageneral}
\ee
and an appropriate Wess-Zumino term. As we will show, the maximal symmetry pattern selects, in all cases, values of $r_1$ and $r_2$ such that this target space reduces to a single complex Grassmannian, which exactly matches the phases $\sigma_{23}$, $\sigma_{43}$, $\sigma_{14}$ of figures \ref{kbig}, \ref{kmedium} and \ref{ksmall}. Interestingly, these values correspond to minimizing the dimension of the target space (\ref{sigmageneral}) with respect to $r_1$ and $r_2$, once we take into account the constraints they obey.

In addition, the scalar condensation leads to a partial or total Higgsing of the gauge group following the pattern
\be
U(n) \longrightarrow U(n-r_1-r_2) ~.
\ee

In the same spirit of appendix \ref{append1}, one can show that the on-shell potential as a function of $(r_1,r_2)$ is never minimized inside the region defined in (\ref{constraint}), so that the minimum of the potential is achieved at the boundaries of this region. The maximal degeneracy of the eigenvalues imply that, under our assumptions, there is never a case in which a sigma model coexists with a TQFT in a given phase. In addition, all the other excitations get a mass either by Higgs mechanism or from the scalar potential, and hence can be safely integrated out.

In order to find the values of $(r_1,r_2)$ and the pattern of Higgsing and global symmetry breaking, one should consider four qualitatively different cases, depending on the value of $n$, see figures \ref{S1}, \ref{S2}, \ref{S3} and \ref{S4}. Tables collect all data necessary to pinpoint the phase the scalar theory enjoys on the $(M^{\,2}_1,M^{\, 2}_2)$ plane, which can be either a TQFT or a sigma model.  

\begin{figure}[t]
\centering
\includegraphics[scale=0.5]{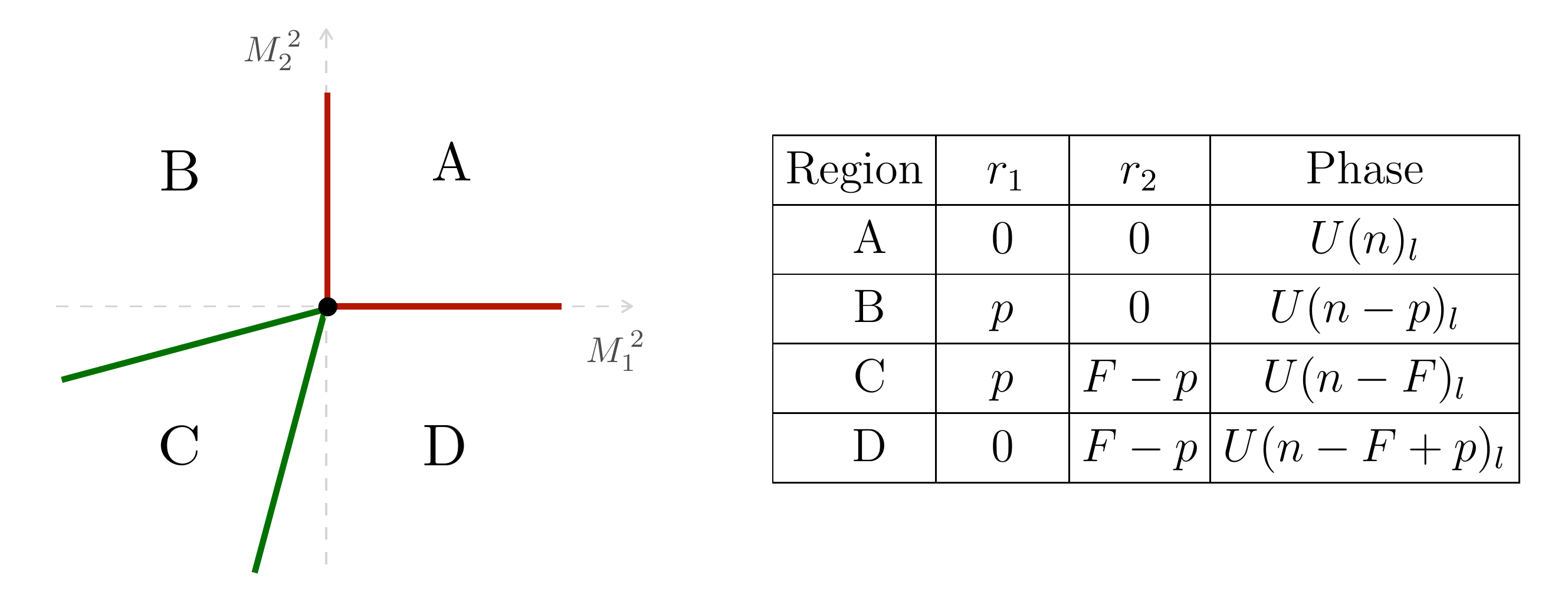}
	\caption{\small{Phase diagram of the bosonic theory in the case $p\leq F-p \leq F\leq n$. The region $\cal R$ of double condensation coincides with $C$.}}
\label{S1}
\end{figure}

\begin{figure}[t]
\centering
\includegraphics[scale=0.5]{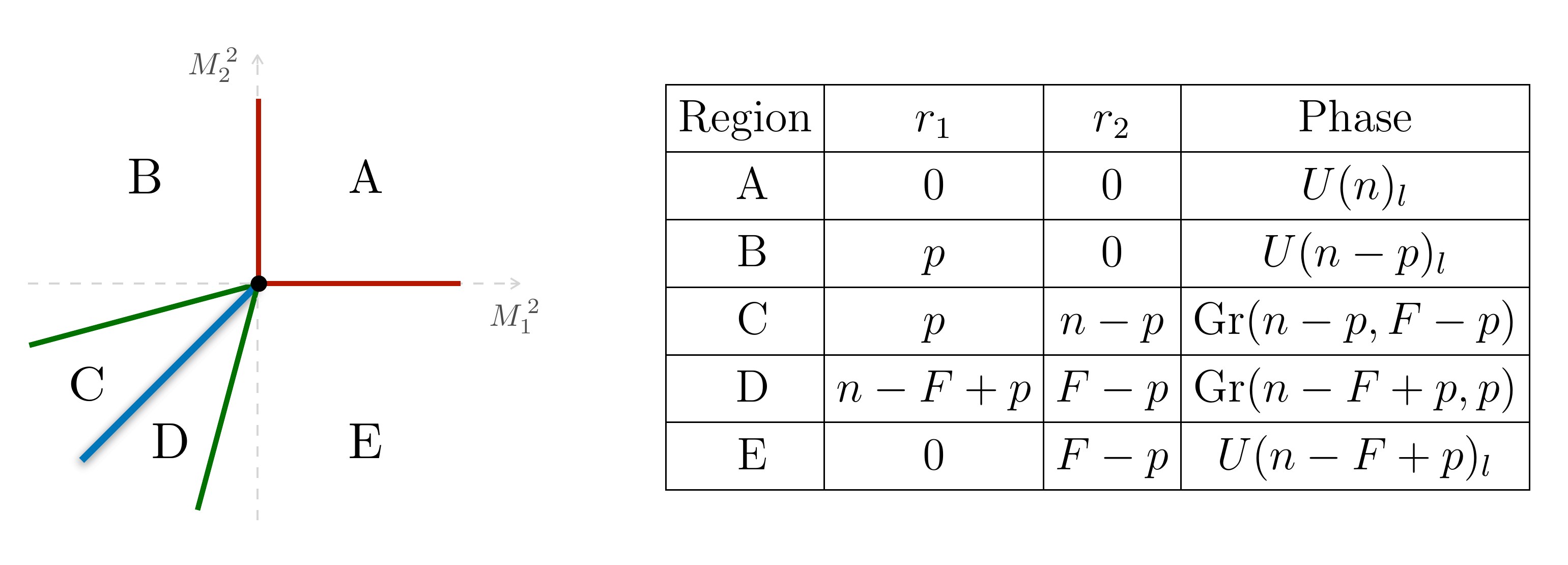}
	\caption{\small{Phase diagram of the bosonic theory in the case $p\leq F-p\leq n<F$. The region $\cal R$ of double condensation coincides with $C+D$, including the blue line.}}
\label{S2}
\end{figure}

\begin{figure}[t]
\centering
\includegraphics[scale=0.5]{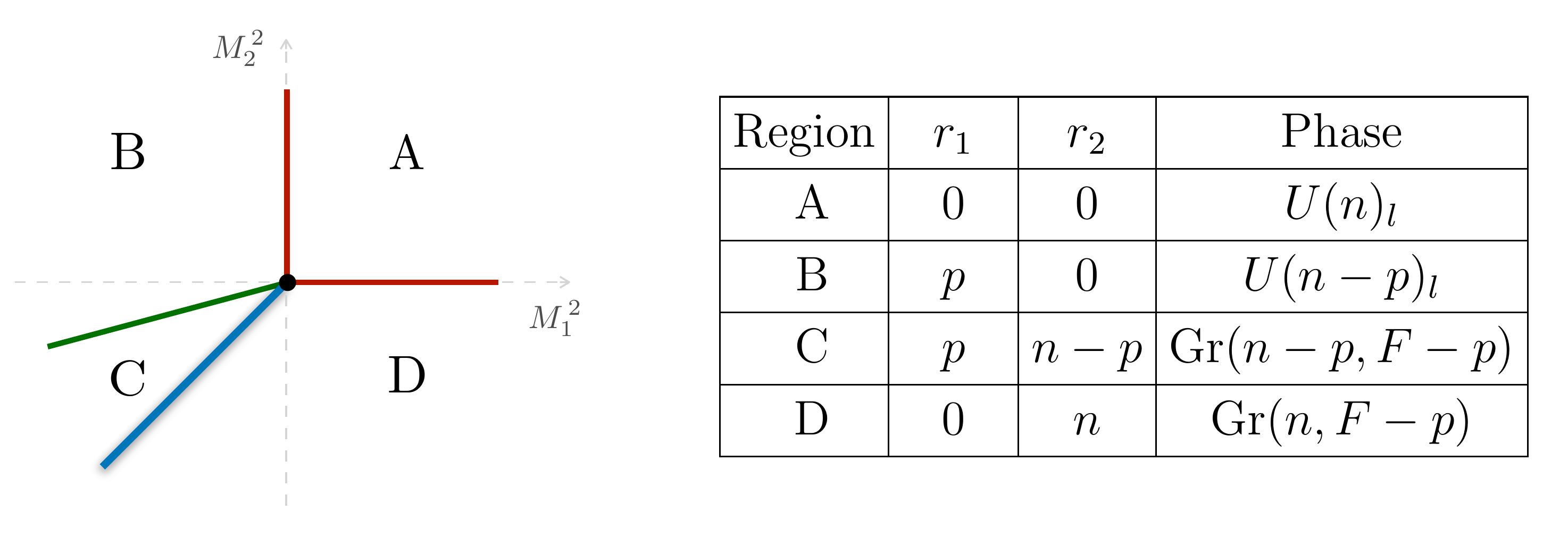}
	\caption{\small{Phase diagram of the bosonic theory in the case $p\leq n<F-p\leq F$. The region $\cal R$ of double condensation coincides with $C$, including the blue line.}}
\label{S3}
\end{figure}

\begin{figure}[t]
\centering
\includegraphics[scale=0.5]{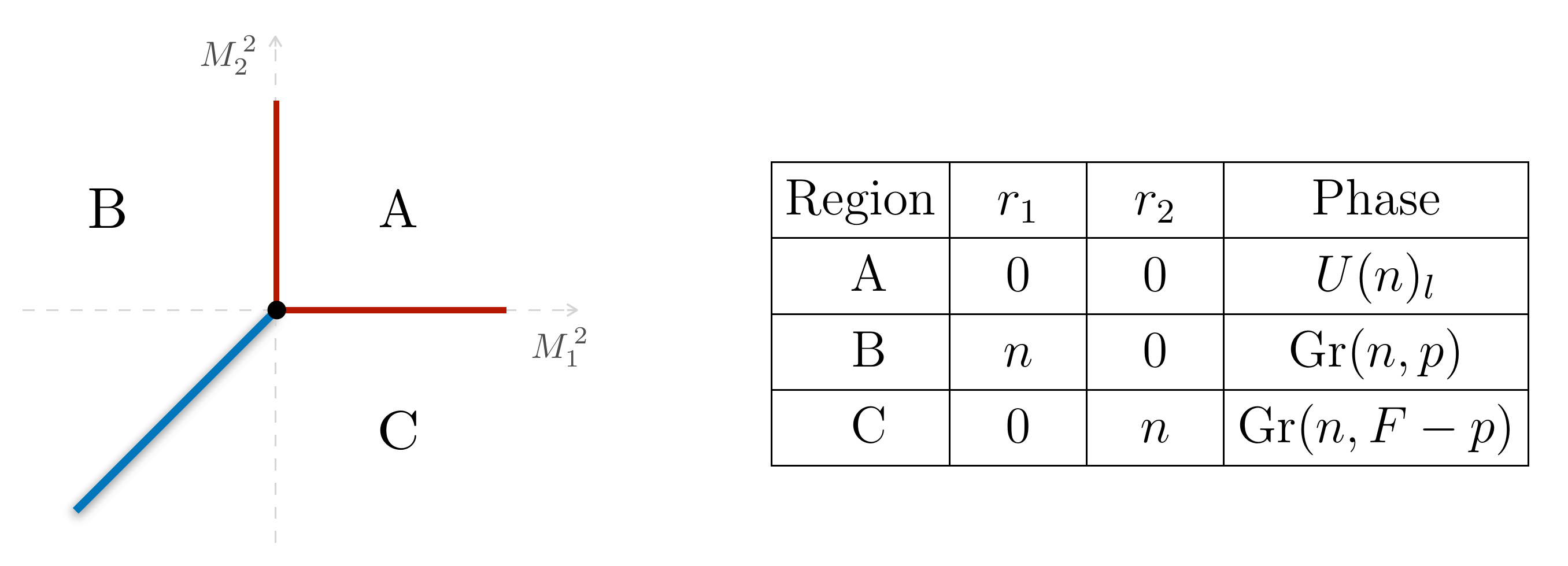}
	\caption{\small{Phase diagram of the bosonic theory in the case $n<p\leq F-p\leq F$. In this case, the region $\cal R$ of double condensation shrinks to the blue line only.}}
\label{S4}
\end{figure}

Starting from the first quadrant, region A, where all scalars have positive mass, red and green lines represent the critical theories where one of the two sets becomes massless and condenses, whereas the blue line is the quantum phase of one-family QCD$_3$. On the red lines the first set of scalars condenses, partially or totally Higgsing the gauge group. In the former case, the components of the second set of scalars charged under the unbroken gauge group may still become massless and condense, this locus corresponding to the green lines in the figure. Neutral components receive instead an additional positive contribution to their squared mass from quartic terms of the potential. One can easily see that this contribution makes their squared mass always positive, so that they never condense and can be integrated out in the whole phase space. When  the gauge group is completely Higgsed by the first condensation, instead, all scalars that have not condensed first cannot give rise to any other critical line.

Let us now specify the values of $n$ to make contact with our conjecture and explore the topological structure of the diagrams around the critical points.

If $n=F/2+k$, the allowed diagrams are given in figures \ref{S1}, \ref{S2} and \ref{S3}. It is now easy to check that the range of validity of each diagram and its various phases exactly reproduce the topological structure of the fermionic diagrams in figures \ref{kbig}, \ref{kmedium} and \ref{ksmall}, respectively, in the neighborhood of the black dot in the first quadrant.

If $n=F/2-k$, the allowed diagrams are given in figures \ref{S3} and \ref{S4}. It is again easy to check that the range of validity of each diagram and its various phases exactly reproduce the topological structure of the fermionic diagrams in figures \ref{ksmall} and \ref{kmedium}, respectively, in the neighborhood of the black dot in the third quadrant. Note that in this case, as in the one-family case, cf figure \ref{KS2}, the orientation of the bosonic diagrams should be reversed.

To summarize, we have shown that the two-dimensional phase diagrams of the bosonic theories are perfectly consistent with the fermionic ones, in the regions of the phase diagrams where the duality is supposed to hold. In particular, we have reproduced the peculiar structure with critical points where more than three critical lines meet.

\subsection{Perturbing $\sigma$ via massive deformations}
\label{saucestain}

In the previous subsection we have shown that the vacuum structure of the fermionic theory is exactly reproduced by the corresponding bosonic dual theories near each transition point. We now want to see what happens when we perturb the non-linear sigma model $\sigma$ (the blue line in figures \ref{kmedium} and \ref{ksmall}) with a small mass term which explicitly breaks the $U(F)$ symmetry to $U(p)\times U(F-p)$.

A similar symmetry-breaking deformation was considered in \cite{Komargodski:2017keh} to check consistency under flowing down from $F$ to $F-1$ ({\it i.e.} a flow in the space of theories). Our philosophy, here, is to choose a symmetry-breaking perturbation that does not change the theory but allows us to investigate the planar region in a neighborhood of the quantum phase $\sigma$.

To do that, we deform the mass of the $p$ scalars $\phi_1$ with a small perturbation $\delta M^{\, 2}$. If $\delta M^{\, 2}>0\ (<0)$ we are investigating the region $M^{\, 2}_2<M^{\, 2}_1\ (M^{\, 2}_2>M^{\, 2}_1)$ where  the set $\phi_2$ ($\phi_1$) condenses first. In fermionic language $M^{\,2}_2<M^{\, 2}_1$ corresponds to $m_2<m_1$ ({\it i.e.} below the $m_1=m_2$ line) for the bosonic theory $U(F/2+k)$ and $m_2>m_1$ ({\it i.e.} above the $m_1=m_2$ line) for the bosonic theory $U(F/2-k)$. Viceversa, for the case $M^{\, 2}_2>M^{\, 2}_1$.

Let us call $n$ the rank of the bosonic theory gauge group, which can be either $F/2+k$ or $F/2-k$. In these conventions, the target space of the sigma-model phase $\sigma$ reads
\be
\mbox{Gr}(n,F)=\frac{U(F)}{U(n)\times U(F-n)} ~.
\ee
Massive deformations act modifying the target space, but they do not change the coefficient $N$ of the Wess-Zumino term. Let us now analyze the different possibilities, performing our analysis in the underlying gauged linear model.

\begin{itemize}

\item If $\delta M^2>0$ and $F-p>n$, the $(F-p)\ \phi_2$ condense first and completely Higgs the gauge group $U(n)$, whereas the $p\ \phi_1$ do not play any role. Indeed, since there is no more gauge group, all the surviving $\phi_1$ are neutral and can be safely integrated out. The resulting sigma model has target space $\mbox{Gr}(n,F-p)$.

\item If $\delta M^2>0$ and $F-p<n$ then the $(F-p)\ \phi_2$ cannot Higgs completely the gauge group, but as they condense they Higgs it down to $U(n-F+p)$ and then can be integrated out. The charged components of the $p\ \phi_1$ then condense, whereas the neutral ones have a positive squared mass around this configuration. This leads to a sigma model with target space $\mbox{Gr}(F-n,p)$.

\item If $\delta M^2<0$ and $p<n$, then the $p\ \phi_1$ condense first and Higgs the gauge group to $U(n-p)$. By the same mechanism as before, the $(F-p)\ \phi_2$ condense, eventually, leading to a sigma model with target space $\mbox{Gr}(F-n,F-p)$.

\item If $\delta M^2<0$ and $p>n$ then the $p\ \phi_1$ condense first and completely Higgs the gauge group, whereas the $(F-p)\ \phi_2$ do not play any role. This leads to a sigma model with target space $\mbox{Gr}(n,p)$.

\end{itemize} 

It is a tedious but simple exercise to check that specifying the above analysis to the case of interest, {\it i.e.}  $n=F/2 \pm k$, the resulting sigma models coincide with those living in the shaded regions around the quantum phase $\sigma$ of phase diagrams in figures \ref{kmedium} and \ref{ksmall}.

It is worth noticing that the above check is not entirely independent from the discussion of the previous section, since we are actually using the underlying gauged linear sigma model. It would be nice to have a proof directly in non-linear sigma-model terms. While the above results would not change, at least qualitatively, such an analysis might shed light on the nature of the phase transition around $\sigma$.

Note that adopting the same philosophy of \cite{Komargodski:2017keh}, instead, one can consistently flow from the theory coupled to $F$ flavors to the one coupled to $F-1$ flavors. This can be done by giving a large mass to, say, one of the $p$ fermions $\psi_1$. After this deformation we get the same duality with parameters
\be
(N,k,F,p) \longrightarrow (N,k\pm 1/2, F-1,p-1) \, ,
\ee
for a positive (negative) mass deformation. 

Using the same approach one can also play with $k$. As already observed in section \ref{kappa0}, our proposal for $k=0$ and $p=F/2$ is consistent with the Vafa-Witten theorem which holds on the entire $m_2=-m_1$ line, figure \ref{k0pF2}. We can use massive deformations to increase the value of $k$, and show that if our proposal is correct for $k=0$, it remains true even for $k>0$. In particular, all values of $k$ up to $F/2-p$ can be reached by integrating out by mass deformations the $p$ fields in the first set, whereas bigger values of $k$, up to $F/2$, are reached by acting on the second set. This shows that one can consistently flow to $(N,k,F,p)$ for any $k\leq F/2$.


\section {Comments and outlook}
\label{conc}

In  this work we have constructed the phase diagram of two-family QCD$_3$, extending the analysis carried out in \cite{Komargodski:2017keh} for the degenerate mass case. While our results agree with \cite{Komargodski:2017keh} in the limits where the two-dimensional phase diagram becomes effectively one-dimensional, there exist  ranges in the parameter space which present novel phenomena. These are inherent to the $m_1 \not = m_2$ case, {\it e.g.} the shaded regions in figures \ref{kmedium}, \ref{ksmall} and \ref{k0gen}, which describe new gapless phases, and phase transitions between them along $\sigma$. 

We now  want to discuss a few directions along which our work could be extended.

As far as the bosonic analysis is concerned, we have limited ourselves to quartic couplings. This is done in analogy to ordinary Wilson-Fisher fixed points for $O(N)$ vector models, and it has been, in fact, the general approach when considering boson/fermion dualities. However, the scalar theory one is dealing with is a gauged $U(N)$ linear sigma model and, as of today, a full understanding of the nature of its fixed points (if any) has not yet been achieved. Strictly speaking, one cannot exclude that along the RG-flow higher order operators acquire large (negative) anomalous dimensions and the effective low-energy theory should take several such operators into account. What one usually does is to start considering those operators whose dimensions near the Gaussian fixed point are the lowest, {\it i.e.} quadratic and quartic couplings. In three space-time dimensions sextic scalar operators are classically marginal, so including them in the analysis would be the first natural extension one should look for. It is possible that the inclusion of such operators does not change qualitatively the picture we have outlined, at least in some region of such an extended space of couplings, but this is a point worth investigating further. 
In a similar vein, one could consider quartic couplings not respecting the full $U(F)$ global symmetry but just $U(p) \times U(F-p)$. This could correspond to yet other relevant deformations of the massless theory, different from mass terms. 
	
If boson/fermion duality is correct, all these (putative) novel relevant deformations should have a counterpart on the fermionic side of the duality (the first natural guess being Gross-Neveu-Yukawa couplings, in analogy with \cite{Metlitski:2016dht}). In this respect, the two-family QCD$_3$ case could work as the simplest laboratory to extend (and check) boson/fermion dualities beyond present understanding. }

Our work does not address the issue of the actual order of the phase transitions in QCD$_3$. While it is known that in certain limits  such phase transitions are second order \cite{Appelquist:1988sr,Appelquist:1989tc,Aharony:2011jz,Giombi:2011kc,Aharony:2012nh,GurAri:2012is,Aharony:2012ns,Jain:2013py,Jain:2013gza}, there are hints this is not the case, in general {\cite{  Armoni:2019lgb}. Having a clear picture about this aspect would give crucial insights on how we should think about boson/fermion dualities in general.

In our two-dimensional phase diagrams, there are some phase transitions that might be more amenable to treatment, namely the transitions between different sigma models. It would be nice to have a more detailed description of these transitions directly in non-linear sigma-model  terms.

Finally, one could explore other situations where  two-dimensional phase diagrams are expected. For instance, situations where matter fermions are in other representations of the gauge group, such as the adjoint. In such cases, two-dimensional phase diagrams can be used as a tool to guess what happens along the diagonal, where the global symmetry is enhanced.

\section*{Acknowledgments} 
We thank Giulio Bonelli, Andres Collinucci, Lorenzo Di Pietro, Frank Ferrari, Alberto Mariotti, Arash Ranjbar and Marco Serone for discussions, and in particular Francesco Benini, Zohar Komargodski and Paolo Spezzati for very useful remarks and comments. We are also grateful to the authors of \cite{Armoni:2019lgb}  for sharing their draft before submission and for correspondence. 
R.A. and P.N. acknowledge support by IISN-Belgium (convention 4.4503.15) and by the F.R.S.-FNRS under the ``Excellence of Science" EOS be.h project n.~30820817, M.B. and F.M. by the MIUR PRIN Contract 2015 MP2CX4 ``Non-perturbative Aspects Of Gauge Theories And Strings" and by INFN Iniziativa Specifica ST\&FI. R.A. is a Research Director and P.N. is a Research Fellow of the F.R.S.-FNRS (Belgium). R.A. and P.N. warmly thank SISSA's Theoretical Particle Physics group for the amazing hospitality during the preparation of this work.


\appendix

\section{$U(n)$ gauge theories coupled to $f$ scalars}
\label{append1}
Let us consider a three-dimensional gauge theory with gauge group $U(n)$ and Chern-Simons level $l$, coupled to $f$ scalar fields in the fundamental representation $\phi_i^{\,\alpha}$, where $\alpha=1,...,n$ is the gauge index and $i=1,...,f$ the flavor index. 

The gauge invariant operator which can be built out of  the scalar fields is the meson field $X=\phi\phi^{\dagger}$. In components
\be
X_{i}^{\,j}=\phi_i^\alpha\phi^{* j}_{\,\alpha} ~,
\ee
which is an $f \times f$ Hermitian matrix whose rank $r$ satisfies
\be
r \leq \mbox{min}(n,f) ~.
\ee
This matrix can be diagonalized and put in the form
\be
X=\mbox{diag}(x_1,x_2,...,x_r,0,...,0) \,,
\ee
where $x_i$ are the $r$ positive eigenvalues of $X$.

The potential which preserves the $U(f)$ global symmetry reads, up to quartic terms in the scalar fields
\be
V=\mu (\mbox{Tr}X) + \lambda (\mbox{Tr}X)^2 + \widetilde\lambda (\mbox{Tr}X^2) \,,
\ee
where we take $\widetilde\lambda>0$, which requires $\widetilde\lambda+  \mbox{min}(n,f) \lambda>0$ in order to make the potential bounded from below. If $\mu \geq 0$ then the minimum of the potential is achieved for $X=0$. This corresponds to the case where the gauge group is not Higgsed, so that at low energies the (massive) scalars can be integrated out and the effective theory is pure $U(n)$ at level $l$. If $\mu < 0$, instead, minimizing the potential one gets the following  equation for the eigenvalues $x_i$
\be
\mu + 2\lambda \mbox{Tr}X + 2\widetilde\lambda \, x_i = 0 ~, 
\ee
which implies that
\be
x_i=\frac{-\mu}{2\lambda r + 2\widetilde\lambda} \quad \forall i=1,...,r ~,
\ee
meaning that all non-vanishing eigenvalues are degenerate. Moreover, on these minima the potential is
\be
V=-\frac{\mu^2 r}{4\lambda r + 4 \widetilde\lambda} ~,
\ee
which is minimized when $r$ is maximum, {\it i.e.} when $r=\mbox{min}(n,f)$. Note that the condition for the eigenvalues $x_i$ being positive is the same which assures the stability of the potential. 

If $f<n$ this means that the Higgsing is maximal, the gauge group is broken in $f$ independent directions and the global symmetry $U(f)$ is unbroken. After integrating out the massive fluctuations of the scalars around their minimum configuration, we get as the resulting IR theory pure $U(n-f)$ at level $l$.

If $f>n$ the gauge group is completely Higgsed and the global symmetry is spontaneously broken to $U(n)\times U(f-n)$, leading to an IR dynamics described by a non-linear sigma model with target space
\be
\mbox{Gr}(n,f) = \frac{U(f)}{U(n)\times U(f-n)} ~,
\ee 
and a Wess-Zumino term with coefficient $|l|$.

Note that the above result, {\it i.e.} maximal Higgsing for $f<n$ and degeneracy of non-vanishing eigenvalues for $f>n$,  depends crucially on assuming a quartic scalar potential with a single trace contribution (similar observations were done, in a different context, in  \cite{Metlitski:2016dht,Hansen:2017pwe} and also appear in \cite{Armoni:2019lgb}). One  should also assume $\widetilde\lambda$ to be positive to get the aforementioned pattern. Indeed, for negative $\widetilde\lambda$ the minimum of the potential (whose stability now requires $\lambda+\widetilde{\lambda}>0$) is achieved at $r=1$, meaning that only one scalar field condenses, giving rise to a different vacuum structure in the negative squared mass phase. 

If one allows higher-order terms in the potential, one would expect that there still exists a region in such a larger space of couplings for which the above extremization pattern holds.  
Clearly, it is not fully satisfactory that one is required to make assumptions on the scalar potential in order to match the fermionic phase diagram. However, gaining complete control on the structure of the scalar potential for generic values of $n,f$ and $l$, in such a non-supersymmetric context, is presently beyond reach.

When applied to QCD$_3$ with one species of fermions, this explains, upon use of boson/fermion duality, the level/rank dualities in the $m<0$ regime, as well as the structure of the Grassmannian  \eqref{sigma}, see figures \ref{KS1} and \ref{KS2}. 

As discussed in section \ref{bosondual}, allowing a scalar potential with all possible gauge invariant operators only up to quartic terms has the same effects as those discussed here also in the more intricate two-dimensional phase space.     


\end{document}